\documentclass[preprint,12pt]{article}
\usepackage{amssymb,times}
\makeatletter
\usepackage{graphicx,subfigure,pstricks}

\def\@cite#1#2{\nolinebreak$^{\scriptstyle #1\if@tempswa , #2\fi}$}
\def\@citex[#1]#2{\if@filesw\immediate\write\@auxout{\string\citation{#2}}\fi
  \def\@citea{}\@cite{\@for\@citeb:=#2\do
    {\@citea\def\@citea{,\penalty\@m}\@ifundefined
       {b@\@citeb}{{\bf ?}\@warning
       {Citation `\@citeb' on page \thepage \space undefined}}%
{\csname b@\@citeb\endcsname}}}{#1}}

\begin{document}

\newcommand{\ea}{et al.}
\newcommand{\be}{\begin{equation}}
\newcommand{\ee}{\end{equation}}

\newcommand{\vr}{\textbf{r}}
\newcommand{\vR}{\textbf{R}}
\newcommand{\vL}{\textbf{L}}
\newcommand{\vK}{\textbf{K}}
\newcommand{\vJ}{\textbf{J}}
\newcommand{\vN}{\textbf{N}}
\newcommand{\vM}{\textbf{M}}
\newcommand{\vX}{\textbf{X}}
\newcommand{\vP}{\textbf{P}}
\newcommand{\vs}{\textbf{s}}
\newcommand{\vx}{\textbf{x}}

\baselineskip 22pt
\parindent 12pt
\parskip 10pt

\noindent {\large\bf Kepler unbound: some elegant curiosities of classical mechanics}\\[0.1in]
\noindent Niall J. MacKay and Sam Salour\\
\noindent {\em Department of Mathematics, University of York, UK}

\vskip 0.2in
\parbox[t]{5.5in}{{\bf ABSTRACT}: We describe two exotic systems of classical mechanics: the McIntosh-Cisneros-Zwanziger (`MICZ') Kepler system, of motion of a charged particle in the presence of a modified dyon; and Gibbons and Manton's description of the slow motion of well-separated solitonic (`BPS') monopoles using Taub NUT space. Each system is characterized by the conservation of a Laplace-Runge-Lenz vector, and we use elementary vector techniques to show that each obeys a subtly different variation on Kepler's three laws for the Newton/Coulomb two-body problem, including a new modified Kepler third law for BPS monopoles.}

\vskip 40pt
\noindent{\large\bf 1. Introduction}

\noindent Newtonian gravity is one of the joys of the undergraduate syllabus, and the first great triumph of western mathematical science. The motion of a planet about the sun is first reduced to a single motion about the centre of mass (the `reduced two-body problem'). When the motion is bound (we call this the `Newton Kepler problem'), orbits are ellipses, and obey Kepler's laws. The problem of the classical motion of electrically-charged particles (the `Kepler-Coulomb problem') is essentially the same---although we know that
quantum mechanics will supervene.

Thus two of the classic problems of physical motion (the orbit of a planet about the sun, and the orbit of electrons about an atomic nucleus) have a special, simple mathematical structure. The underlying reason is that such problems have the maximum number of conserved quantities possible for 3D motion (and in fact this is true of a third classic problem, the harmonic oscillator, although we shall not treat this here). In 3D, six initial conditions (a position and a momentum) determine motion completely, so a maximum of five conserved quantities is possible, with one dimension (the position on the ellipse) remaining.\cite{int} Four of these quantities are furnished by energy and angular momentum, intuitively familiar to all physicists. But the fifth is the one independent component of the Laplace-Runge-Lenz (LRL) vector,\cite{G} which provides a neat short-cut to the geometrical equation of the ellipse, by-passing the usual solution of the equation of motion. Unlike energy and angular momentum, the LRL vector is not straighforwardly associated via Noether's theorem with a natural geometrical symmetry of the Kepler problem.\cite{Noether}

The LRL vector is nonetheless moderately well-known, and its generalizations appear in many contexts.\cite{GenLRL} In these (`superintegrable') systems, Poisson brackets of conserved quantities are typically  polynomial, rather than linear, in the conserved quantities.\cite{LV} For the LRL vector the brackets are quadratic. Happily, when the energy, which commutes with the other conserved quantities, is fixed, the symmetry becomes a conventional Lie algebra: $so(3,1)$ (for positive-energy, scattering states) or $so(4)$ (for negative-energy, bound states). It is in this, algebraic form, applicable to quantum mechanics, that the LRL vector, famously, may be used to find the states of the hydrogen atom.\cite{H}

Our story is a simpler one, of elementary classical mechanics. It is principally about two problems, both originating in the study of magnetic monopoles (which we shall call the `MICZ Kepler problem'\cite{MICZ} and  the `Taub NUT Kepler problem'\cite{M,GM}). Both of these rather exotic systems have LRL vectors, and each realizes a subtly different variation of Kepler's laws, with
some beautiful attendant geometry. Both can be embedded within more advanced mathematics, but our
central point is that one can see the full beauty of the Kepler problems using no more than elementary vector techniques: `dot' and `cross' products and their derivatives, applied to central and Lorentz forces.

Section two sets the scene by recalling the role of the LRL vector in the Newton Kepler problem. Section three treats the MICZ Kepler problem, in which the most striking variation is that the field source is not at the focus of the ellipse but rather at the apex of the cone whose intersection with the orbital plane is the elliptical orbit. Kepler's second and third laws here apply to radial vectors on the surface of the cone, rather than in the orbital plane. Section four treats the Taub NUT Kepler problem, in which the slow motion of two well-separated solitonic monopoles again results in elliptical motion. Kepler's second law here applies in the orbital plane, but the third law is now subtly different.

\vskip 10pt
\noindent{\large\bf 2. The Newton Kepler problem}

\noindent Recall the reduced two-body problem in Newtonian mechanics.\cite{Goldstein} One begins with two bodies, of masses $m_1,m_2$ at positions $\vx_1,\vx_2$, and acting on each other with force of magnitude $Gm_1m_2/r^2$, where $G$ is Newton's constant,  $\vr:=\vx_1-\vx_2$ and $r=|\vr|$. When external forces on the bodies are proportional to their masses (as, for instance, in a uniform external gravitational field), the separation vector $\vr$ obeys
\be  \ddot\vr = \ddot\vx_1-\ddot\vx_2=\left({1\over m_1}+{1\over m_2}\right) {Gm_1m_2 \over r^2}\hat \vr\,,\ee
or
\be \mu \ddot\vr =  {Gm_1m_2 \over r^2}\hat \vr\,, \ee
where $\mu$ is the `reduced mass' $m_1m_2/(m_1+m_2)$ and  $\vr=r\hat\vr$. This simplifies to
\be \ddot\vr = - {GM \over r^2}\hat\vr\,\label{E1}\ee
where $M$ is the total mass $M=m_1+m_2$. Conservation of angular momentum and energy, $$\vL=\vr \times \dot\vr\quad {\rm and} \quad E= {1\over 2} \dot\vr\cdot \dot\vr - {GM \over r}$$ (both per unit of reduced mass), follows.

Notice that, in situations where one body is much more massive than the other ($m_2\gg m_1$, say), such as a small planet orbiting a star, we have $M\simeq m_2$ and $\mu\simeq m_1$, while $\vx_2$ is approximately the centre of mass.

The standard method for arriving at an equation for the orbit begins with the observation that, because $\vL$ is constant, the motion is planar. Using polar coordinates $r,\theta$ for the orbital plane, we have $L=r^2\dot\theta$  and the radial component of (\ref{E1}) is
\be
\ddot{r} - r\dot{\theta}^2=\ddot{r} - {L^2 \over r^3} = -{GM\over r^2}\,.\label{E2}\ee
Writing $u=1/r$ we have
\be \dot{r}=-L{ du\over d\theta}\,,\qquad \ddot{r}=-L^2 u^2 {d^2u \over d\theta^2}\,,\ee
which allows us to eliminate the explicit $t$-dependence and obtain from (\ref{E2}) a purely geometrical equation relating $u=1/r$ to $\theta$:
\be
{d^2 u\over d\theta^2} + u = {GM\over L^2}\,.
\ee
The solution is the conic section
\be
{1\over r} = {GM\over L^2}\left( 1+e\cos\theta\right)\,, \quad{\rm where} \quad
e=\sqrt{ 1 +{2EL^2\over G^2M^2}}\,.\label{E5}
\ee
This gives a nice connection between the geometrical and physical parameters. The length-scale is $L^2/GM$, and the orbit  is bounded when $E<0$ and $e<1$. For $E>0$ we have $e>1$ and we see from (\ref{E5}) that $r\rightarrow\infty$ for some $\theta$, leading to hyperbolic scattering. Finally $E=0,e=1$ is the parabolic case.

An alternative route is to note that the LRL vector
\be
{\vK}={\dot\vr}\times{\vL} - GM\hat{\vr}
\ee
is also conserved, for which one needs to know that
\be\label{unitr}
{d\over dt}\hat{\vr} = {{\vr}\times (\dot{\vr}\times {\vr})\over r^3}\,.
\ee
Then we immediately have that
\be
\vK \cdot\vr = Kr\cos\theta=L^2-GMr\,,
\ee
or
\be
 {1\over r} = {GM\over L^2} \left( 1 + {K\over GM} \cos\theta\right)\,.
\ee
The seven apparent components of $E,\vL$ and $\vK$ are reduced to five---the largest number possible---by the
relations $\vK\cdot\vL=0$ and $K^2=G^2M^2+2EL^2$, which in turn can be used to derive the second equation of (\ref{E5}).\cite{Egan}

Kepler's laws for this problem are: first, that orbits are ellipses, here derived (\ref{E5}) entirely within the orbital plane; secondly, that the radial vector sweeps out equal areas in equal times, which follows from conservation of $L$; and, thirdly, that the ratio of the square of the orbit's period to the cube of its major axis is the same for all closed orbits. This is at its simplest for circular orbits, where equating the inverse-square and centripetal accelerations (\ref{E1}) at angular speed $\Omega$ gives $r\Omega^2=GM/r^2$. More generally, one computes from the second law and the area of the ellipse that
\be
 {a^3 \over T^2} = {GM\over 4\pi^2} \,,\label{K3}
\ee
where $a$ is the semi-major axis and $T$ is the period.

\vskip 10pt
\noindent{\large\bf 3. The MICZ Kepler problem}

\noindent Consider now the motion of a charged particle in the presence of a magnetic charge (a `monopole') fixed at the origin.\cite{monopole} Under the Lorentz force, the particle's position $\vr$ obeys
\be\label{monopoleeom}
\ddot\vr = \alpha {\dot\vr \times \vr \over r^3}\,, \label{F1}
\ee
where $\alpha$ is proportional to the product of the monopole's magnetic charge and the particle's electric charge. The particle's speed $v=\vert \dot\vr \vert $ is conserved,
\be
{d\over dt} v^2 =2 \dot\vr\cdot\ddot\vr = 2 \alpha  {\dot\vr\cdot \dot\vr \times \vr \over r^3}=0\,,
\ee
and its distance $r$ from the origin obeys
\be
{d^2\over dt^2} r^2 = 2{d\over dt} \vr\cdot\dot\vr = 2 v^2\,.\label{F2}
\ee
It has a conserved, modified angular momentum $\vJ=\vL-\alpha \hat\vr$ (where $\vL=\vr\times \dot\vr$ as in Sect.\ 2, and one uses (\ref{unitr},\ref{monopoleeom}) to prove conservation), which satisfies $J^2=L^2+\alpha^2$ (so that $L$ is still conserved even though $\vL$ is not). Further,
\be
\vJ\cdot \vr = -\alpha r\,,\label{cone}
\ee
so that the particle moves on a cone, of opening angle $2\theta$ where
\be\theta=\cos^{-1} (\alpha/J)=\cos^{-1}\left(1+L^2/\alpha^2\right)^{-1/2}\,.\ee
Thus the greater the particle's angular momentum $L$, the more open the cone.
If the particle does not reach the origin, it spirals inwards on the cone at constant speed, reaches a minimum distance from the origin, and spirals out again.

The motion remains confined to the cone if we add a central force to (\ref{F1}). In particular, suppose we add a Coulomb force, so that the charged particle is now moving in the presence of a `dyon' of mixed electric and magnetic charge.\cite{dual} Then
\be
\ddot\vr = \alpha {\dot\vr \times \vr \over r^3} - \beta {\vr \over r^3}\,, \label{F3}
\ee
where $\beta$ is proportional to the product of the two electric charges. In contrast to the pure monopole case, the particle's speed is no longer conserved, while (\ref{F2}) becomes
\be
{d^2\over dt^2} r^2 = 2{d\over dt} \vr\cdot\dot\vr = 2 \left( v^2 -  {\beta \over r}\right)\,.\label{E3}
\ee
Thus circular motion is possible only if $v^2r=\beta$ (a simple case of the virial theorem).

The MICZ Kepler problem\cite{MICZ} adds another term to the equation of motion, a fixed inverse-{\em cube} force, which stabilizes the motion on the cone.\cite{BO} Suppose we have
\be
\ddot\vr = \alpha {\dot\vr \times \vr \over r^3} - \beta {\vr \over r^3} + \gamma {\vr \over r^4}\,. \label{F4}
\ee
Then $\vJ$ is still conserved (since the $\beta$ and $\gamma$ terms are both central), and the particle still moves on the cone. The conserved energy is
\be
E={1\over 2} v^2 - {\beta \over r} + {\gamma \over 2r^2}\,.
\ee
The key result is the existence of a LRL vector
\be
\vK=\dot\vr\times\vJ - \beta\hat\vr\,,
\ee
which is conserved if and only if $\gamma=\alpha^2$, a condition which is applied throughout the remainder of this section.\cite{Lecht} Then $\vK\cdot\vJ=\alpha\beta$ and $K^2=2L^2E+\beta^2$, so that we again have five independent constants of the motion, as required. Further
\be
\vr\cdot\vN=\alpha L^2\,,\label{plane}
\ee
which defines a plane, normal to $\vN=\alpha\vK-\beta\vJ$. Combining (\ref{cone},\ref{plane}) we see that the particle therefore moves in the intersection of a cone with a plane, a conic section. After having computed $\vN\cdot\vJ=-\beta L^2$ and $N^2=L^2(2\alpha^2E+\beta^2)$, we can compare the angle between $\vN$ and $\vJ$ with the cone angle to find that the section is an ellipse when $E<0$, and a circle when $E=-{\beta^2 \over 2 J^2}$. It is rather satisfying thus to derive the conic section directly, rather than arriving at the ellipse entirely within the orbital plane as in Sect.\ 2. Note that, in contrast to the Newton Kepler problem, the field source is at the apex of the cone rather than in the orbital plane.

We can use conservation of $\vJ$ to rewrite the equation of motion (\ref{F3}) as
\be
\ddot\vr = - {\alpha \vJ + \beta \vr \over r^3}\,.
\ee
Observing that $-\alpha\vJ/\beta$ is in the orbital plane and defining $\vR = \vr + \alpha\vJ/\beta$, we have
\be
\ddot\vR = -\beta {\vR \over r^3}
\ee
and we observe that the motion is under a central force in the orbital plane---but one whose centre, at $-\alpha\vJ/\beta$, is on the axis of the cone and so is {\em not} a focus of the orbital ellipse,\cite{dandelin} and which does not vary as the inverse-square of the distance from the centre.

The way to make sense of this rather peculiar situation is to perform what we call the {\em cone anthesis} (Figure 1), which opens out the surface of the cone: a vector on the cone is rotated, within the plane containing it and the cone axis, onto the (`horizontal') plane containing the apex of the cone and normal to its axis.\cite{anthesis} The transformation is $\vr\mapsto\vX$, where
\be
\vX = {\vJ \times (\vr \times \vJ) \over JL}={\widehat\vJ \times (\vr \times \widehat\vJ) \over \sin\theta}\,.
\ee
(Note that one could replace $\vr$ by $\vR$ in this formula, since they differ by a vector parallel to $\vJ$.) One then has $r=X$ and thus
\be
\ddot\vX = -{\beta \over X^2} \widehat\vX\,,
\ee
a Newton Kepler problem in the horizontal plane, whose solutions are ellipses.\cite{line} Kepler's second and third laws thus also hold in this plane, so that by inverting the anthesis we see that Kepler's second law applies to the area swept out by $\vr$ on the surface of the cone, and the third law to distances measured on the cone or in the horizontal plane, rather than in the orbital plane.

\pagebreak

\begin{pspicture}(-7,-4)(10,13)

\psellipse(0,10)(5,1.3)
\psline(0,0)(-5,10)
\psline(0,0)(5,10)
\psline(0,0)(0,13)

\psellipse[linewidth=2pt, rot=20](0.46,5.9)(3.1,1)
\psellipse[linewidth=2pt, linestyle=dashed, rot=15, dash=4pt 2pt](1.28,0.52)(6.3,2.55)


\rput{-70}{\psellipticarc[linestyle=dashed, dash=1pt 2pt]{<-}(0,0)(6.5,5.5){67}{160}}

\psline(-8,-4)(7,-4)
\psline(-4,4)(8.5,4)
\psline(-4,4)(-8,-4)
\psline(7,-4)(8.5,4)

\psline[linewidth=2pt]{->}(0,0)(2.08,5.54)
\psline[linewidth=2pt]{->}(0,0)(5.6,-0.3)
\psline[linewidth=2pt]{->}(0,0)(0,5.8)

\put(-0.3,-0.47){$O$}
\put(5.65,-0.7){$X$}
\put(1.9,5.75){$R$}
\put(-0.4,5.7){$P$}

\end{pspicture}

\vspace*{0.5in}
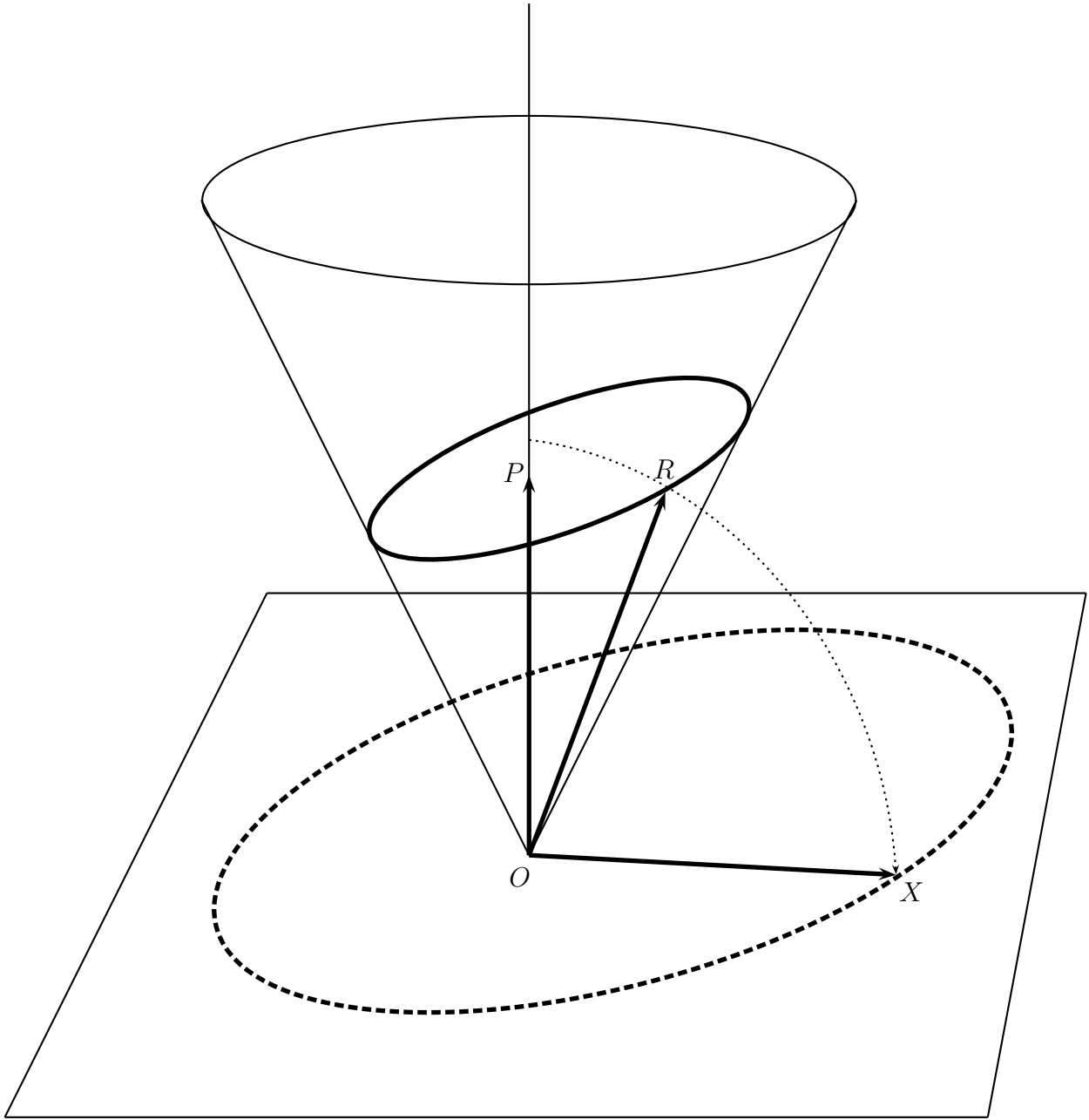
\centerline{\parbox[t]{5.7in}{Figure 1: the {\bf cone anthesis}. The elliptical orbit (bold) is the intersection of the cone with the orbital plane (not shown). The cone anthesis (lightly dotted arrow) opens the cone out by rotating the orbital vector $\vr=\overrightarrow{OR}$, within the plane spanned by it and the cone axis $\overrightarrow{OP}$, onto the horizontal plane (which contains $O$ and is perpendicular to the cone axis), where $\vX=\overrightarrow{OX}$ traces a new ellipse (bold dashed). In the orbital plane itself, where $-\alpha\vJ/\beta=\overrightarrow{OP}$, the vector $\vR=\overrightarrow{PR}$  moves under a central force about $P$, but this is not an inverse-square law in $|PR|$, and $P$ is not a focus of the orbital ellipse. In the horizontal plane, however, the motion is that of the Newton Kepler problem.}}

\vskip 10pt
\noindent{\large\bf 4. The Taub NUT Kepler problem}

\noindent Magnetic monopoles are not observed in our world,\cite{real} still less the exotic field of (\ref{F4}). But they have a significant theoretical implication which {\em is} observed, namely that electric charge is quantized. A slick demonstration of this follows from quantization of $\vJ\cdot\hat\vr$, since $\alpha$ is proportional to the product of electric and magnetic charge.\cite{Saha}

In modern gauge field theories, charge quantization follows from the properties of representations of Lie groups. However, such field theories, where they incorporate electromagnetism, in fact contain magnetic charges in the form of solitary waves (`solitons'), known as BPS monopoles.\cite{Olive} The adiabatic ({\em i.e.\ }reversible) motion of pairs of these is described by geodesic motion on their parameter (or `moduli') space, whose metric is known. In the limit of large separation, this becomes geodesic motion on euclidean Taub NUT space---which again turns out to be a Kepler problem.\cite{M,GM}

For dyonic monopoles of unit magnetic charge and relative electric charge $Q$, the equation of motion is
\be
\left(1-{2\over r}\right) \ddot\vr = \left( \vert\dot\vr\vert^2-Q^2\right){\vr \over r^3} +2Q {\dot\vr \times \vr \over r^3} - 2 {(\vr\cdot\dot\vr)\dot\vr \over r^3}\,,
\ee
in which the unit length is inversely proportional to the monopole mass\cite{M,GM}.
This equation conserves
\begin{eqnarray}
E & = & {1\over 2} \left(1-{2\over r}\right)\left( \vert\dot\vr\vert^2+Q^2\right)\\
\vJ & = & \left(1-{2\over r}\right) \vr \times \dot\vr - 2Q \hat\vr\\
\vK & = & \left(1-{2\over r}\right) \dot\vr \times \vJ + 2(E-Q^2)\hat\vr\,,
\end{eqnarray}
which satisfy $E>0$, $J^2=L^2+4Q^2$ (so that $L=\vert (1-2/r)\vr\times \dot\vr\vert$ is conserved), $K^2= L^2(2E-Q^2)+4(E-Q^2)^2$ and $\vJ\cdot\vK=4Q(Q^2-E)$. Thus, as in Sect.\ 2, there are five independent conserved quantities.\cite{KK}

Defining $\vN=Q\vK + (E-Q^2)\vJ$ (of length $N=LE$), we find that
\begin{eqnarray}
\vJ\cdot\hat\vr = -2Q && - {\rm \;a \;\; cone}\\
\vN\cdot\vr = QL^2 && - {\rm \;a \;\; plane}
\end{eqnarray}
 and we again have a conic section. The opening angle of the cone is $2\nu$ where
 \be
 \nu= \cos^{-1}(2Q/J) = \cos^{-1} \left( 1 + {L^2\over 4Q^2} \right)^{-1/2}\,,
 \ee
and is thus fixed by $L$. The angle between the cone axis and the normal to the plane is
\be
\cos^{-1} \left( {Q^2/E-1 \over \sqrt{1+4Q^2/L^2}} \right)\,,
\ee
fixed by $E$. The distance of the plane from the origin is then $QL/E$, so that, on a cone of given opening angle, there is precisely one elliptical orbit of given eccentricity, confirming Gibbons and Manton's expectation ([9], p205).

Further, the equation of motion may be rewritten as
\be
\ddot\vr = -{2E \over (r-2)^3} \left( \vr - {\vK \over E} + {Q \vJ \over E}\right)\,,
\ee
so that, upon defining $\vP=(\vK-Q\vJ)/E$ and $\vs=\vr-\vP$, we have
\be
\ddot\vs = -{2E \over (r-2)^3} \vs\,.
\ee
(We may check that $\vP\cdot\vN=QL^2$, so that $\vP$ is in the orbital plane.) Finally, after computing $\vP\cdot\vr=2r + L^2/E$ and $P^2=4+2L^2/E$, we find that $s^2=\vert\vr-\vP\vert^2=(r-2)^2$, so that
\be
\ddot\vs = -{2E\over s^2}\hat\vs\,.
\ee

Thus, astonishingly, we have a Kepler problem in the orbital plane. Note that (in contrast to the MICZ Kepler problem) this is expressed via a simple inverse-square law in the orbital plane, and Kepler's first and second laws apply there without modification---even though the origin of the separation vector $\vr$ is at the apex of the cone. Unlike in the previous sections, the central force is now proportional to $E$, so that Kepler's third law no longer holds in its standard form. However, by replacing $GM$ with $2E$ in the Newton Kepler problem and considering the motion at perihelion $s=l/(1+e)$, we find that
\be
2E = {s \over s+2} \left( v^2 + Q^2\right) = {2E l \over s(s+2)} + {s\over s+2}Q^2
\ee
and thence, applying $l=a(1-e^2)$, we obtain a modified third law
\be
4\pi^2 {a^3\over T^2} = 2E = { a Q^2 \over a+1 }\,, \qquad {\rm or} \qquad {a^2(a+1) \over T^2} = {Q^2 \over 4\pi^2}\,,
\label{K3'}
\ee
in which the relationship between semi-major axis and period remains both independent of eccentricity $e$ and, as $a$ becomes large, approximately (to use Newton's term) sesquialterate.

For a geometrical equation for the orbit we first define a vector in the orbital plane,
\be
\vM = \vP - {Q\over E-Q^2} \vJ = {1\over E}\vK + {Q(2E-Q^2) \over E(Q^2-E)}\vJ \,.
\ee
We can then calculate
\be
\vM \cdot \vs = \left( \vP +{Q\over Q^2-E}\vJ \right)\cdot\left(\vr-\vP\right)\,.
\ee
Writing $\vM\cdot\vs = MS\cos\varphi$, we find that
\be
{1\over s} = {2E\over L^2} \left( 1 + {M(Q^2-E) \over 2E}\cos\varphi \right)\,,
\ee
where the eccentricity
\be
e= {M(Q^2-E) \over 2E} = \left( 1 +{L^2\over 2E} - {Q^2L^2\over 4E^2} \right)^{1/2}
\ee
so that, for a closed orbit $e<1$, we must have $E<Q^2/2$.

These results specialize straightforwardly to the results for circular orbits found by Gibbons and Manton,\cite{GM} in their equations (5.21-5.23). For such an orbit we require that $\vM=0$ (effectively, that $\vJ$ and $\vK$ be collinear), and thus that
\be
M^2={4E^2+2EL^2-L^2Q^2 \over (E-Q^2)^2}=0\,, \label{M2}
\ee
so that in a circular orbit $E$ is determined by $L$.
In such an orbit of constant $s$ and speed $v=s\Omega$, we have that $L={s\over r}rv=sv$ and $E={s\over 2r}(v^2+Q^2)$, reproducing (5.21,5.23).  The modified Kepler third law is that
\be
s^3\Omega^2=2E={s\over r} \left( s^2\Omega^2 + Q^2 \right)\,,
\ee
so that $\Omega (r-2)\sqrt{r-1}=Q$  (as may also be derived by setting $e=0$ in (\ref{K3'})), which is (5.22) of [9].

\pagebreak
\noindent{\large\bf 5. Concluding remarks}

\noindent All of these systems can be placed within a unified mathematical framework.\cite{unity} The broader context is that of superintegrability, where Bertrand's theorem and the classification of superintegrable systems\cite{Evans} lead on to deep algebraic structures (``finite $W$-algebras") which have been the subject of much recent research by both mathematicians and physicists.\cite{fW} Another strand of research extends the MICZ Kepler problem to higher dimensions.\cite{Meng}

Physically, it seems that the origins of the Laplace-Runge-Lenz vector may lie in a problem's (special) relativistic invariance\cite{Dahl}---so that, for example, the hidden symmetry of the Taub NUT Kepler problem would originally be due to the Lorentz covariance of Yang-Mills-Higgs theory.\cite{Gibbons} So there may well be a direct line of ascent from Kepler's laws to relativity!

The results described in this article are therefore rather more than merely the elegant curiosities of our title. Integrability---exact solvability---is, to paraphrase Wigner, unreasonably ubiquitous in theoretical physics. In an age dominated by complexity and simulation, it seems quaintly old-fashioned to focus on problems that can be solved by exact, pencil-and-paper methods. Yet integrable systems have enjoyed a resurgence over the last generation, culminating in the discovery of their central role in the (`AdS/CFT') correspondence between string theory and gauge theory. This may yet yield the first-ever exact (to all orders in $\hbar$) solution of a quantum field theory.\cite{Beisert}
The return to significance of the kind of ideas that underlie the Kepler problem would not have surprised Newton, whose scientific manifesto was of ``nature ... very consonant and conformable to her self ... and very simple".\cite{Newton}

The examples we have presented are not among the empirical phenomena of (at least this part of) our world, but they are among the few examples of classical integrability which can be accessed using elementary techniques, and their geometry is strikingly elegant. We hope that they have given the reader at least a flavour of the beauty of the subject.

\vskip 10pt
{\bf Acknowledgments}

\noindent We should like to thank Ed Corrigan for comments.



\begin{thebibliography}{99}
\bibitem{Gibbons}
We should like to thank Gary Gibbons for this observation.
\end{thebibliography}
\end{document}